\documentstyle[twoside,11pt,aaspp4]{article}
\tighten
\eqsecnum
\received{}
\accepted{}
\journalid{}{}
\articleid{}{}
\slugcomment{To appear in {\it The Astrophysical Journal}}

\begin{document}

\title{New rotation periods in the Pleiades: Interpreting activity
indicators} 

\author{Anita Krishnamurthi\altaffilmark{1},\altaffilmark{2}, D. M.
Terndrup\altaffilmark{1}, M. H. Pinsonneault and K. Sellgren}
\affil{Department of Astronomy, The Ohio State University, 
174 W. 18th Ave., Columbus, OH 43210 \\
Electronic mail:
anita,terndrup,pinsono,sellgren@astronomy.ohio-state.edu}  

\author {John R. Stauffer and Rudolph Schild}
\affil{Smithsonian Astrophysical Observatory, 60 Garden St.,
Cambridge, MA 02138\\ Electronic mail: stauffer@cfa.harvard.edu,
rschild@cfa.harvard.edu}

\author {D. E. Backman,  K. B. Beisser, D. B. Dahari, A. Dasgupta, J. T.
Hagelgans and M. A. Seeds}
\affil{Dept. of Physics and Astronomy, Franklin and Marshall
College, P.O. Box 3003, Lancaster, PA 17604 \\ 
Electronic mail: dana@astro.fandm.edu, m\_seeds@acad.fandm.edu}

\author{Rajan Anand, Bentley D. Laaksonen, Laurence A. Marschall and T.
Ramseyer}
\affil{Dept. of Physics, Gettysburg College, Gettysburg, PA
17325\\
Electronic mail: marschal@gettysburg.edu, ramseyer@goethe.uca.edu}

\altaffiltext{1}{Visiting Astronomer, Kitt Peak National Observatory, 
National Optical Astronomy Observatories, which is operated by the
Association of Universities for Research in Astronomy, Inc. (AURA) under
cooperative agreement with the National Science Foundation.}
\altaffiltext{2}{Current address: JILA, Univ. of Colorado, Campus Box
440, Boulder, CO 80309}

\begin{abstract}

We present results of photometric monitoring campaigns of G, K and M
dwarfs in the Pleiades carried out in 1994, 1995 and 1996.  We have
determined rotation periods for 18 stars in this cluster.  In this
paper, we examine the validity of using observables such as X-ray
activity and amplitude of photometric variations as indicators of
angular momentum loss.  We report the discovery of cool, slow rotators
with high amplitudes of variation.  This contradicts previous
conclusions about the use of amplitudes as an alternate diagnostic of
the saturation of angular momentum loss.  We show that the X-ray data
can be used as observational indicators of mass-dependent saturation in
the angular momentum loss proposed on theoretical grounds.

{\it Subject headings :} stars: activity -- stars: evolution -- stars:
rotation -- stars: spots -- X-rays: stars

\end{abstract}

\section{Introduction}

Angular momentum evolution in stars is intimately related to the stellar
magnetic field because protostars are thought to be magnetically coupled
to circumstellar accretion disks and also because stars lose angular
momentum in magnetized winds.  The magnetic field is related to stellar
activity, including coronal activity such as X-ray emission,
chromospheric activity such as H$\alpha$ and Ca II emission, and
photometric variability due to star spots.  Unfortunately, it is
generally true that the stellar winds from low mass stars, the angular
momentum loss, and even the magnetic field strength and structure are
not directly observed.  Observational studies of angular momentum
evolution have therefore centered on correlations between stellar
activity as indicators of magnetic activity, and other stellar
properties such as mass, age and rotation rate.

The observed distributions of rotation velocities in open clusters of
different ages reveal that the rate at which stars lose angular momentum
on the main sequence depends both on their age and their mass (e.g.,
\cite{sta91b}).  The angular momentum loss rate, $\dot{J}$, is
proportional to $\omega B^{2}$ for low to moderate values of $\omega$,
where $\omega$ is the angular velocity and $B$ is the mean surface
magnetic field (\cite{weber67}).  As $B$ is proportional to $\omega$ for
a linear dynamo, $\dot{J}$ is proportional to $\omega^{3}$ for slow
rotators.  If this is also true for high values of $\omega$, then rapid
rotation would be suppressed prior to the main sequence as fast spinners
would undergo heavy angular momentum loss (\cite{pkd90}).  The
observation of rapid rotators on the main sequence in young open
clusters such as $\alpha$ Per (50 Myr) and the Pleiades (70 Myr)
(\cite{sta84b}; \cite{sta85}; \cite{sta94a}) thus requires a saturation
of $\dot{J}$ at high rotation rates in theories of angular momentum
evolution.  There is also some theoretical support for a decreased
sensitivity of $\dot{J}$ to $\omega$ at higher rotation rates
(\cite{mestel87}; \cite{mac91}).

It has long been known that chromospheric activity and X-ray activity
are correlated with rotation among lower main-sequence stars, gradually
increasing from the slow rotators to the rapid rotators and then
reaching a plateau (\cite{noyes84}; \cite{ros85}).  This plateau is
sometimes interpreted as due to a saturation in the mean surface
magnetic field.  The amplitude of brightness variations in low-mass
stars has also been proposed as an alternate indicator of the saturation
of the surface magnetic field (\cite{odell95}).

There is considerable scatter in plots of stellar chromospheric and
coronal activity vs. rotation which is greatly reduced if activity is
plotted vs. the Rossby number, defined as the ratio of the rotation
period ($P_{rot}$) to the convective overturn time ($\tau_{conv}$)
(\cite{noyes84}; \cite{simon85}; \cite{patten96}).  As $\tau_{conv}$
depends on mass, this suggests that stellar activity in stars depends
both on rotation and on mass (e.g., \cite{patten96}).

Reproducing the mass-dependent rapid rotation and spindown has been a
challenge for theoretical models until recently (see Krishnamurthi et
al. 1997 for a review of theoretical efforts).  Some models, using a
constant value of the saturation threshold that depends only on the
angular velocity, cannot reproduce the mass dependence of the rapid
rotator phenomenon (e.g., Kawaler 1987, 1988, Chaboyer et al. 1995a,
1995b) while others suppress rapid rotation prior to the main sequence
(e.g., \cite{pkd90}).

Some recent models (\cite{cam94}, \cite{barnes96}) have considered a
mass-dependent angular momentum loss law to explain the observed
distribution of rotation velocities of open cluster stars.  The models
of Collier Cameron and Li (1994) have their starting point on the main
sequence.  They use an explicit mass dependence in the scaling constant
in the angular momentum loss and a very high mass-dependent saturation
threshold for $\dot{J}$ (45-75 $\omega_{\odot}$).  However, given these
parameters, Krishnamurthi et al. (1997) find that rapid rotation is
suppressed prior to the main sequence for solar analogs with saturation
thresholds greater than 20 $\omega_{\odot}$.  Barnes \& Sofia (1996)
also note that different masses require different saturation thresholds
in order to account for the ultra fast rotators in the young cluster
data.  However, they use an empirical fit for $\omega_{crit}$ as a
function of mass.  Recent models of angular momentum evolution by
Krishnamurthi et al.  (1997) successfully reproduce the observed
distribution of rotation rates in open clusters of different ages by
assuming the angular momentum loss saturation depends on the Rossby
number (and hence mass), not angular velocity alone.  As it is known
that the same mass dependence reduces the scatter in the $P_{\rm
rot}$-X-ray plane (\cite{patten96}), this establishes a connection
between theory and observations that we explore further.

In this paper, we report new period measurements for 18 stars in the
Pleiades, increasing the number of known rotation periods in this
cluster to 51.  The correlation between activity and rotation in the
Pleiades and other open clusters has long been studied using $v \sin i$
measurements (e.g., \cite{sta94}).  As there is now a large sample of
rotation periods in the Pleiades, the correlation can be studied free of
ambiguities introduced by the unknown angle of inclination.  We use this
larger dataset of rotation periods to explore which measure of stellar
activity is the best observational indicator of stellar magnetic field
strengths and their saturation.

\section {Data Acquisition and Analysis}

As part of an ongoing program to determine the photometric rotation
periods for stars in open clusters, we selected targets in the Pleiades
spanning a range of magnitudes, from $V$=10 ($M\sim 1.0M_{\odot}$) to
$V$=16 ($M\sim 0.4M_{\odot}$), in order to better define the correlation
of activity indicators and rotation over a wide mass range.  We obtained
differential photometry for three sets of targets over three observing
seasons (Oct. 1994-Jan. 1995, Oct. 1995-Jan. 1996, and Oct. 1996), using
the Kitt Peak 0.9 m telescope, the 1.8 m and 1.1 m telescopes at Lowell
Observatory in Flagstaff, the NURO 0.8 m telescope, also in Flagstaff,
and the 1.2 m telescope at Mount Hopkins.  Table 1 summarizes the
characteristics of the various telescope and detector combinations.  We
opted to obtain differential photometry as this technique allowed us to
utilize nights with marginal weather that made uniform absolute
photometry difficult.  This maximized the number of nights that could be
used for observations.  Each of the target fields was imaged in
broadband $V$.

\subsection{NURO data}

The data obtained on the NURO 0.8m telescope was reduced separately as
the field of view was very small (4$\arcmin$).  The first step in the
reduction of the NURO data was to remove the zero-read bias and to
divide by a flatfield image.  Aperture magnitudes were then used to
derive the differential photometry from the reduced images.  The
instrumental magnitudes for the target star and the brightest nearby
stars were derived using the routine APPHOT in the
IRAF$^{1}$\footnotetext[1]{The Image Reduction and Analysis Facility
(IRAF) is distributed by the National Optical Astronomy Observatories,
which is operated by AURA, Inc., under contract to the National Science
Foundation.} package.  The star aperture was set to have a radius of 1.5
times each night's average seeing FWHM, and the sky values measured in a
contiguous annulus with a width of 1.5 FWHM.  The difference in
magnitude between the target and the comparison stars in the
instrumental $V$ band pass was then computed.  These differential
magnitudes were then used to compute the rotation period.

Here and for the observations made on other telescopes, we made no
attempt to transform the magnitudes to a standard system.  It was
sometimes the case with the NURO data, particularly for the brightest
targets, that the nearby comparison stars were much fainter than the
target star.  This meant that the differential magnitude of the target
could not be determined with precision.  For the 1994-1995 and 1995-1996
observing seasons, we typically used the NURO data in the determination
of the rotation period when the uncertainties in the differential
photometry were better than 0.025 mag on average.  Only the NURO group
obtained data during 1996.  The targets selected for observation in 1996
were expected to have short rotation periods based on high $v \sin i$
values and on tentative period estimates for some of these stars from
the 1994-95 observing run.

\subsection{Other data}

The data from the other telescopes were treated differently and we now
describe this procedure.  We began by doing the zero-exposure and flat
field corrections, the latter from a combination of dome and sky flats
when available.  Rather than designating a set of predetermined
comparison stars, we derived instrumental magnitudes for {\it all} stars
on each reduced frame using version 2.0 of DoPhot (\cite{she93}).  We
ignored all stars with central intensities which exceeded the linearity
regime of the detector.  This yielded the differential magnitudes of
each target against all the remaining stars on each frame.

To assemble the photometry for each star into a time series, we first
grouped the DoPhot output files by target star.  Then for each target
star, we first generated a time series for the observations from each
telescope.  The final step was to tie the observations on different
telescopes together by using the conventional method of scaling the
magnitudes against predetermined comparison stars.  We did this by
identifying 2-4 stars located near the target Pleiades member on the CCD
chip with the smallest field of view.  The scatter in the relative
magnitudes of the comparison stars around their mean value gave an
estimate of the accuracy of this process; normally the uncertainty was
0.01 mag or better.  Then the offsets were computed to bring the
comparison stars to the scale defined by the magnitude of the target
star on the zero frame for that series.  This offset was applied to all
the magnitudes in the time series, thus rescaling the several time
series to one common magnitude scale.

\subsection{Determination of periods}

We determined rotation periods using two different techniques.  The
first technique used the ``Period'' algorithm (\cite{press93}) for the
analysis of unevenly sampled time-series data.  This routine also yields
a measure of the false alarm probability (FAP) which indicates the
significance of the derived period.  A small value for the false alarm
probability indicates a small probability that the peak is spurious (i.e
due to random fluctuations).

The second technique used computes the best fitting sine wave to a given
set of data.  This routine utilizes the ``Powell routine'' (as described
in \cite{press93}) and computes the period based on the minimum
chi-squared value.  This routine also weights the data points by their
photometric uncertainties.  The uncertainty in the derived rotation
period was computed to be the standard deviation of a Gaussian fit to
the peak corresponding to the minimum chi-squared value.  We used this
routine to independently confirm our derived period.  Both routines
yielded the same period for the stars reported here.

Table 2 contains a summary of the observations, where the first column
is the name of the star observed, column 2 is the number of observations
per star, column 3 indicates the date/data range of the observations and
column 4 specifies the telescope(s) used.  Table 3 lists some properties
of the stars we observed - column 2 is the $V$ magnitude, column 3 is
the $B-V$ color, column 4 is the $v \sin i$ measurement, if known, and
column 5 is the logarithm of the X-ray luminosity of the stars
normalized to the bolometric luminosity.  Table 4 presents the newly
determined periods; FAP represents the False Alarm Probability.  The
light curves are shown in Figure 1.  We were successful in deriving
periods for 21 of the 36 stars in our total sample.  In addition to the
new periods, we recovered periods for three known rapid rotators,
Hii1883, Hii2244 and Hii2927.  The values of 0.23d, 0.56d and 0.26d,
respectively, agree with the previously determined values (\cite{van82};
\cite{sta87b}).

\section{Indicators of magnetic field saturation}

Chromospheric and coronal emission are often used as proxies for
magnetic field strength.  As mentioned earlier, these activity
indicators are known to be correlated with rotation and are believed to
saturate at some value of rotation.  Armed with a considerably larger
database of rotation periods in the Pleiades, we now re-examine plots of
various activity indicators vs. the rotation period.

O'Dell et al. (1995) examined rotation data for G and K single dwarfs
with 0.55$\leq B-V \leq$1.40.  From a study of the amplitude of
variation of the stellar brightness over one rotation period, $\Delta
V$, they concluded that $\Delta V$ increased with decreasing $P_{\rm
rot}$ beyond the saturation threshold inferred from the chromospheric
activity ($\sim$ 3 days for solar type stars, \cite{vilhu84}).  They
argued that the spot coverage continues to increase with increasing
rotation rate, although the chromospheric activity appears to have
saturated.  They proposed that the amplitude of photometric variation
might be a better indicator of the saturation of the magnetic field than
the chromospheric activity, and derived an upper limit on the period for
saturation at $P_{rot} \sim$ 12 hours.  We examine this by considering
the dataset for the Pleiades, adding our new data to the sample, in the
same color range.

As there are more X-ray data available for stars in open clusters, we
start by plotting the chromospheric (H$\alpha$ and Ca II) versus coronal
(X-ray) activity indicators for stars in the Pleiades in Figure
\ref{actx}.  This demonstrates that these activity indicators are indeed 
correlated with each other.

In Figure \ref{actprot}, we plot chromospheric H$\alpha$ emission, the
Ca II infrared triplet (data from Soderblom et al. 1993b), coronal X-ray
emission (data from \cite{micela90} \& \cite{sta94}) and the amplitude
of variation (compilation of older data (open circles) from O'Dell
(1995); new data (filled circles) from this paper) for stars in the
Pleiades against the rotation period.  The amplitude of variation has
been defined consistently in both datasets as the difference between the
maximum and minimum magnitudes.  The plot shows that the
activity-rotation relations are similar for both chromospheric line
emission and broadband X-ray emission, with an increase in activity
levels towards decreasing $P_{rot}$.  We note here that the trends
for H$\alpha$ and the Ca II lines appear to be have a more linear
correlation with rotation than is seen in the X-ray data.  However, more
X-ray data are available in the Pleiades for stars with measured $P_{\rm
rot}$, than H$\alpha$ and Ca II data.  The trends for all three
indicators look very similar if the dataset for the X-ray emission is
restricted to only those stars with measurements in the other two.  The
additional X-ray data helps to show a ``saturation'' type behaviour more
clearly.

There is no trend seen for $\Delta V$ when our new rotation periods
are included with the previously measured periods.  It is also seen from
the plot of $\Delta V$ against $P_{rot}$ that higher amplitudes of
variation exist at longer periods than seen previously (filled symbols
in the range $P_{rot}$=2-10 days, $\Delta V \geq$ 0.1 mag.).  Thus
$\Delta V$ does not increase with increasing rotation rates beyond the
saturation limit of 3 days derived from chromospheric indicators
(\cite{vilhu84}).

O'Dell et al. (1995) also noted that the coolest stars in their sample
($B-V>$1.15) had low amplitudes ($\Delta V<$0.12 mag) and speculated
that this effect had some physical significance.  In Figure
\ref{bvamp}, we plot amplitude vs. $B-V$ color including the
uncertainty in the amplitude which is due to the uncertainty in the
photometry.  We find that four of the stars for which we have now
measured $P_{rot}$ in the Pleiades are cool stars with high
amplitudes and this result is true even when the uncertainty in the
amplitudes is taken into account.  Thus there does not appear to be a
decrease in $\Delta V$ for cooler stars when the sample size is
increased.  Furthermore, the amplitude appears to vary at different
epochs for a given star.  Hii1883, the fastest rotator found to date in
the Pleiades, has been monitored for over fifteen years.  The amplitude
of this star has been found to vary from 0.04 to 0.20 magnitudes over
this time period (see Figure 13 in \cite{sod93a}).  Because three or
four cycles of brightness modulation are required to be certain of the
derived period, most reported periods for cluster stars are for rapid
rotators ($P_{rot}$ a fraction of a day to a couple of days) as they
only need shorter observing runs to accurately measure a rotation
period.  Few stars with modest rotation periods ($P_{rot}$ of 4 days
and more) have had multiple amplitude measurements.  By contrast, some
of the rapid rotators have been studied over many years, and are
therefore more likely to have light curves from several different epochs
and may have been observed with a high amplitude in one of those epochs.
In O'Dell et al. (1995), only the largest observed amplitude was
selected for analysis.  Hence, we conclude that the use of $\Delta V$ as
an indicator of saturation of the magnetic field is ineffective.


\section{Observational indicators of \.J}

We know that both the chromospheric and coronal activity indicators are
better correlated over a wide mass range with the Rossby number,
$N_{R}$, than with rotation period alone.  Hence, the saturation of the
chromospheric and coronal emission depends both on rotation and mass (as
the convective overturn time is a strong function of mass).  Figure
\ref{actross} illustrates this correlation for the Pleiades where the
chromospheric activity indicators as well as log(L$_{x}$/L$_{bol}$) are
plotted against the inverse Rossby number, $N_{R}^{-1}$, calculated
using theoretical convective overturn times from Kim \& Demarque (1996).
We also plot $\Delta V$ vs. the inverse Rossby number and note that this
does not help to reduce the scatter.  There does not appear to be a
well-defined correlation between $\Delta V$ and $P_{rot}$ or between
$\Delta V$ and Rossby number for the Pleiades dataset.

Figure \ref{actross} shows that $\Delta V$ varies by a factor of $\sim$
6 for $N_{R}^{-1}$=10-100.  This is comparable to the variation in
$\Delta V$ at different epochs (a factor of $\sim$5) found by Soderblom
et al. (1993a) for Hii1883.  As X-ray activity in the Pleiades was
studied by the {\it Einstein Observatory} and then by ROSAT a decade
later, the scatter in X-ray activity with time for a star is observed to
be a factor of at most $\sim$5 (\cite{micela96}).  Figure \ref{actross}
shows that the range in X-ray activity for $N_{R}^{-1}$=10-100 is a
factor of $\sim$40.  Thus, while it is possible that $\Delta V$ depends
on $P_{rot}$ or $N_{R}^{-1}$, the dependence is masked by the
scatter, unlike the X-ray activity.

As shown by Figure \ref{actross}, the X-ray activity is very well
correlated with the inverse Rossby number, plateauing at $N_{R}^{-1}$ of
$\sim$ 20.  In their theoretical models of angular momentum evolution in
solar-type stars, Krishnamurthi et al. (1997) found that a
mass-dependent angular momentum loss rate best reproduced the
distribution of rotation rates seen in open clusters.  The mass
dependence used in their models was precisely this Rossby scaling.
Thus, the Rossby number parameterizes the saturation threshold required
both by theory (to explain the rotation distributions of open clusters)
and by observations (to best describe the X-ray and chromospheric
activity).  Hence, the saturation of X-rays in open cluster stars may be
used as an observational indicator of the saturation of the angular
momentum loss law.  Identifying the precise level of X-ray saturation
and its mass dependence will help to pin down the level at which the
angular momentum loss saturates.  This is crucial and should be a goal
for ground-based and space-based studies to determine rotation periods
and X-ray activity levels.

\section{Conclusions}

We have presented results of our photometric monitoring campaign of G, K
and M dwarfs in the Pleiades.  X-ray emission has long been known to
correlate roughly with rotation period and has been shown to correlate
well with Rossby number (defined as the ratio of the rotation period to
the convective overturn time, hence a function of both rotation and
mass).  We conclude that the mass-dependent saturation of the X-ray
emission is consistent with the mass-dependent saturation of the angular
momentum loss required theoretically.  Thus, we now have an observable
that may be used to determine the threshold for angular momentum loss
saturation.  Additional periods and X-ray data obtained in the turnover
region would aid in determining the functional form of the mass
dependence.  The lack of slowly rotating, cool, high amplitude variable
stars in previous observations had led to some discussion on the use of
the photometric amplitudes as an alternate diagnostic of saturation.
However, our discovery of such stars in the Pleiades shows that the
distribution of the amplitudes does not indicate a different saturation
threshold than that inferred from X-rays.

The mass range of the stars included in our sample ranges from roughly
0.5-1.2 $M_{\odot}$.  There is strong observational evidence that the
mass dependence of the angular momentum loss law continues for lower
mass stars (\cite{sta97}; \cite{jones96}).  Establishing the mass
dependence of the saturation in activity for the lower main sequence may
provide valuable clues for models of stellar winds.  In particular,
there seems to be no obvious change in the rotation properties of
stars at the boundary between stars with radiative cores and those which
are fully convective ($\sim$0.25 $M_{\odot}$).

A Rossby-scaled angular momentum loss law would imply high saturation
thresholds for stars more massive than 1.0 $M_{\odot}$, where the
convective overturn timescale decreases rapidly with increasing mass.  A
loss law of the form used by Krishnamurthi et al. (1997) would suppress
rapid rotation in higher mass stars.  There is little evidence for
angular momentum loss in either the pre-MS or the MS for stars with $M
>$ 1.6 $M_{\odot}$, while there is pre-main-sequence spindown in the
intermediate-mass regime ($M \sim$ 1.3-1.6 $M_{\odot}$, \cite{wolff97}).
The highest mass models (1.2 $M_{\odot}$) considered by Krishnamurthi et
al. (1997) spun down too quickly when compared with the cluster data
when a Rossby scaling for the saturation threshold was adopted.  This
may indicate either a change in the nature of the wind for stars of mass
greater than 1 $M_{\odot}$ or a defect in the particular angular
momentum loss prescription of Krishnamurthi et al. for high saturation
thresholds.

We close by noting that there are alternate means of limiting the
angular momentum loss rates for rapid rotators.  Buzasi (1997) has
proposed a systematic tendency towards polar activity for lower mass
stars as a possible explanation for the mass dependence of angular
momentum loss (see also Solanki, Motamen \& Keppens 1997).  There are
also other effects, for example a tendency towards a more complex
magnetic field geometry at high field strengths, which could act to cap
the angular momentum loss rates for rapid rotators (e.g.,
\cite{mestel84}).  A good observational database of activity and
rotation measurements as a function of mass and time may prove valuable
for evaluating the relative merits of the different classes of
theoretical models.
 
\acknowledgements

Observations were obtained with the Perkins 1.8m reflector of the Ohio
Wesleyan and Ohio State Universities at Lowell Observatory, the 1.1m
Hall telescope at Lowell Observatory, the 1.2m Mt. Hopkins telescope at
Whipple Observatory and at the Lowell Observatory 0.8m telescope, which,
under an agreement with Northern Arizona University and the NURO
Consortium, is operated 60\% of the time as the National Undergraduate
Research Observatory.  We thank the KPNO staff for their technical
support during our runs on the 0.9m as well as R. Mark Wagner and Ray
Bertram at Lowell Observatory for their technical support and expertise.
We thank Jodie Dalton, Greg Piecuch and Phil Turcotte for assistance in
observations at the NURO telescope as well as in reduction of the NURO
data.  We also wish to thank Charles Prosser at CfA for obtaining some
measurements for this project and to Sidney Wolff for providing valuable
comments.  AK wishes to thank S.M. Long for discussions and help with
the period-finding routines.  Franklin and Marshall College and
Gettysburg College thank the University of Delaware/Bartol Research
Institute's NASA Space Grant College consortium for support of NURO
membership and partial support of student travel.  Financial support for
this project was provided by NSF AST-9528227 to DT and MHP, an OSU Seed
Grant to MHP and by NASA LTSA program grant NAGW-2698 to JRS.

We would also like to thank the anonymous referee for constructive
comments which improved the paper significantly.

\clearpage

\begin{figure}
\caption{Differential light curves in $V$ for Pleiades stars.  The
different symbols indicate data taken on different telescopes.  The
crosses are data from the KPNO 0.9m telescope, the open circles are from
the SAO 1.2m telescope, the filled triangles are from the 1.8m Perkins
telescope, the open triangles are from the 1.8m Perkins telescope in
2$\times$2 binned mode, the stars are from the NURO 0.8m telescope and
the squares are from the 1.1m Hall telescope.  The error bars indicate
the photometric uncertainties.}
\end{figure}

\begin{figure}
\caption{Correlation of chromospheric and coronal activity indicators
for available data in the Pleiades.  (top) The ratio of the H$\alpha$
flux to the stellar bolometric flux, R(H$\alpha$), vs. the X-ray
luminosity normalized to the bolometric luminosity,
(L$_{x}$/L$_{bol}$). (bottom) The ratio of the flux in the 8542\AA{ }
line of Ca II to the bolometric flux, R[Ca II(8542)] vs.
(L$_{x}$/L$_{bol}$).  H$\alpha$ and Ca II data from Soderblom et
al. (1993b); X-ray data from Stauffer et al. (1994). }
\label{actx}
\end{figure}

\begin{figure}
\caption{Activity indicators for Pleiades stars vs. rotation period,
$P_{rot}$. (top) Chromospheric H$\alpha$ emission and Ca II infrared
triplet at 8542\AA{ }.  (bottom) The X-ray luminosity normalized to the
bolometric luminosity, L$_{x}$/L$_{bol}$, and the amplitude of variation
at V due to starspot modulation, $\Delta V$.  H$\alpha$, CaII and X-rays
increase with decreasing rotation period.  H$\alpha$ and Ca II data from
Soderblom et al. (1993b); X-ray data from Stauffer et al. (1994).
Rotation period data: {\it Open symbols} - Previously existing periods
from Magnitskii (1987), Stauffer et al. (1987), Prosser {\it et al.}
(1993) and Prosser et al. (1995), as compiled by O'Dell et al. (1995).
{\it Filled symbols} - Our new periods.}
\label{actprot}
\end{figure}

\begin{figure}
\caption
{The amplitude of photometric variation, $\Delta V$, vs. $B-V$ color for
Pleiades stars.  {\it Open symbols}: Periods compiled by O'Dell et
al. (1995). {\it Filled symbols}: Our new periods.  Error bars represent
the uncertainty in the amplitude determination due to the photometric
uncertainties.}
\label{bvamp}
\end{figure}

\begin{figure}
\caption{Activity indicators for Pleiades stars vs. the log of the
inverse Rossby number, $P_{rot}$/$\tau_{conv}$, where $\tau_{conv}$
is the convective overturn timescale.  (top) Chromospheric H$\alpha$
emission and Ca II infrared triplet at 8542\AA , (bottom)
(L$_{x}$/L$_{bol}$) and $\Delta V$.  Symbols represent same as in
Fig. \ref{actprot}.}
\label{actross}
\end{figure}

\clearpage

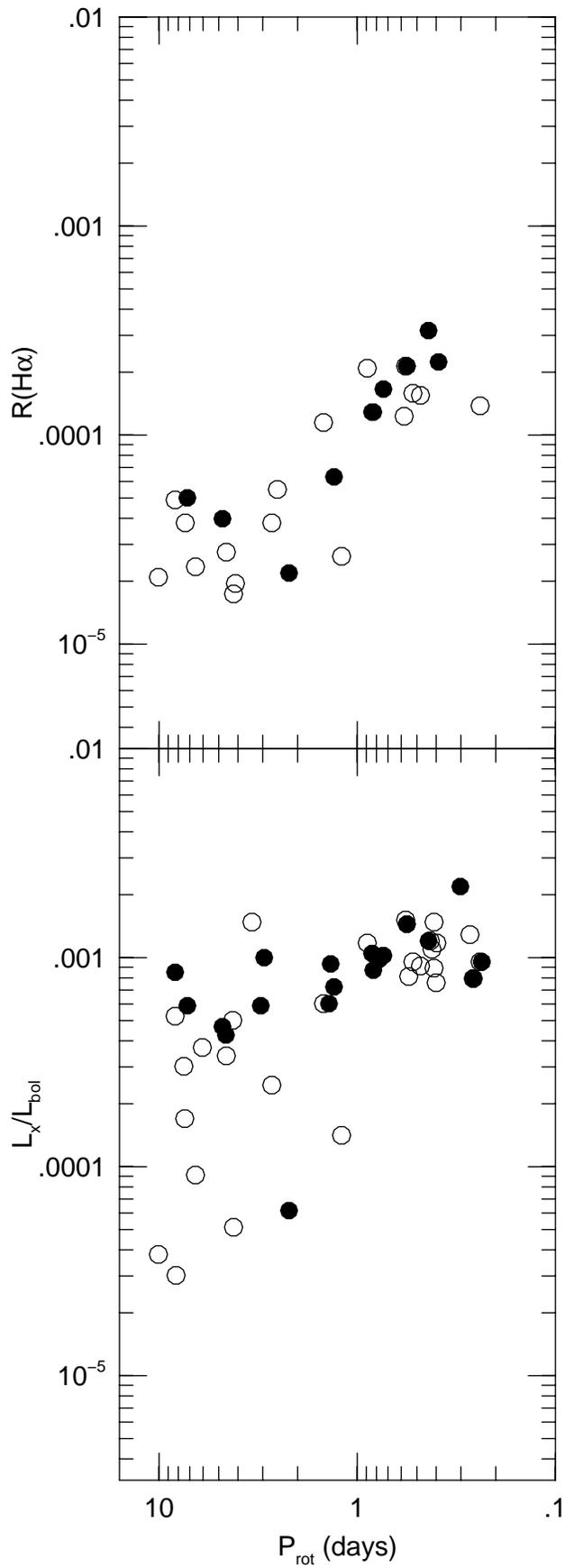
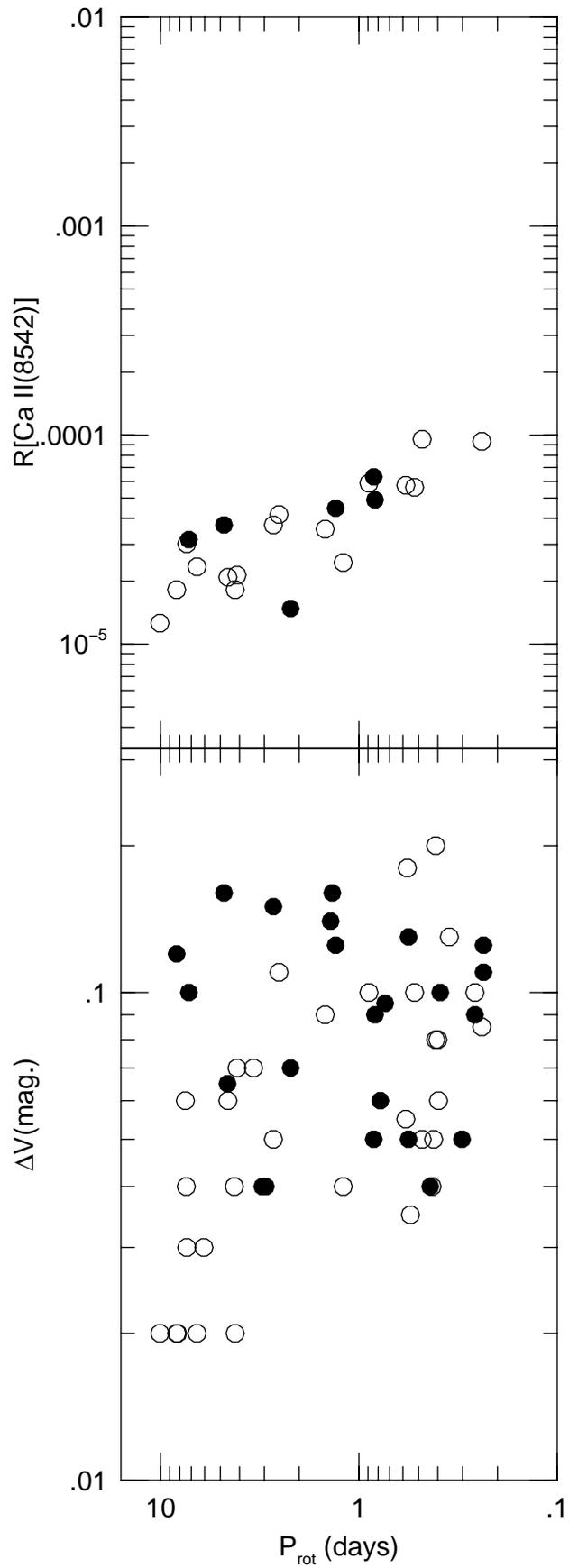
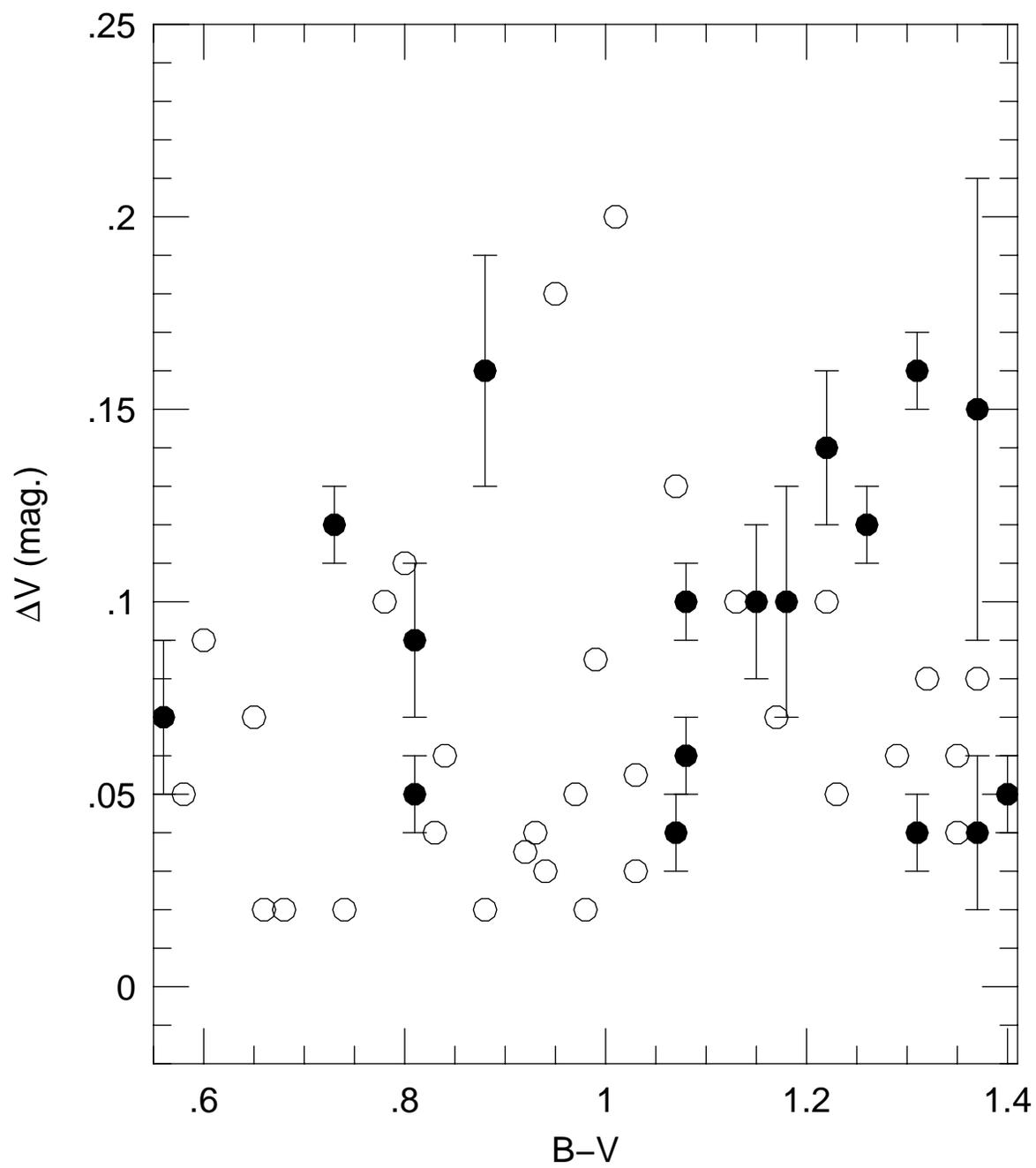
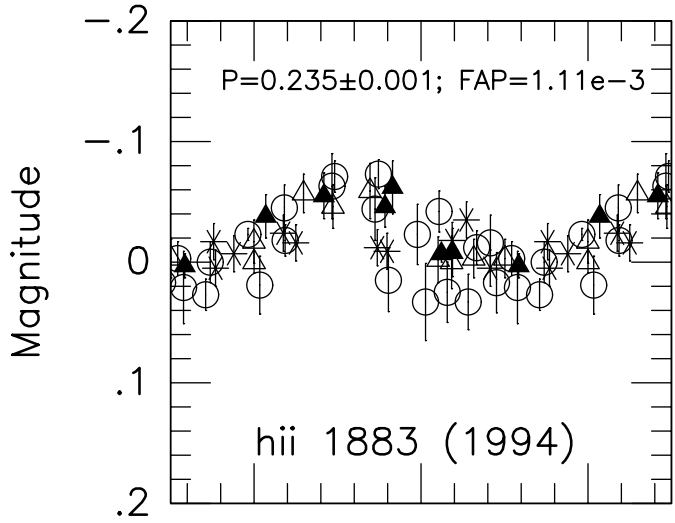
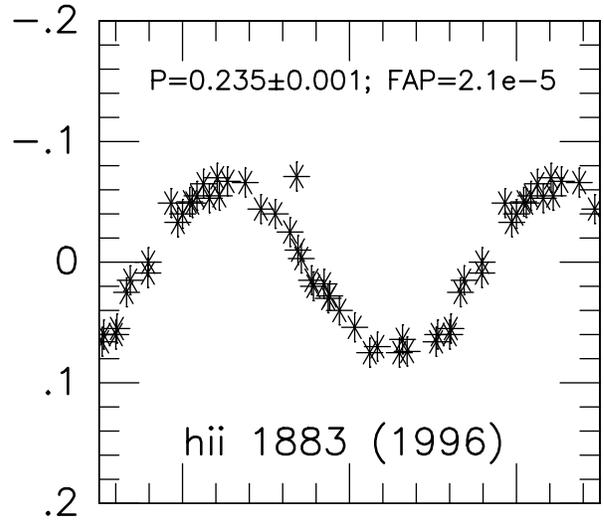
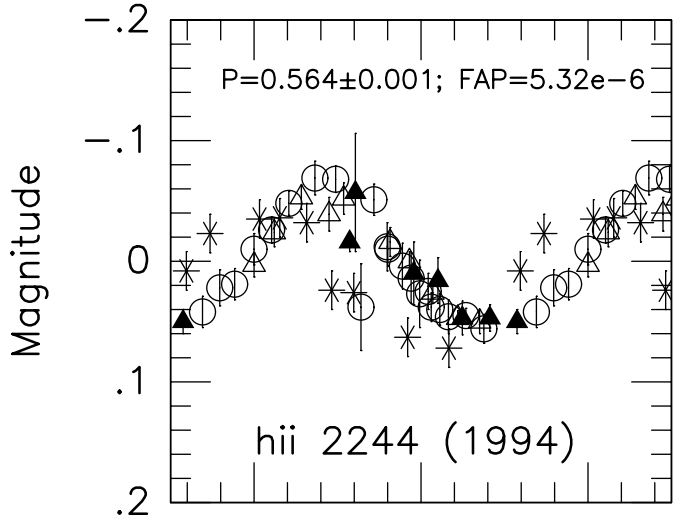
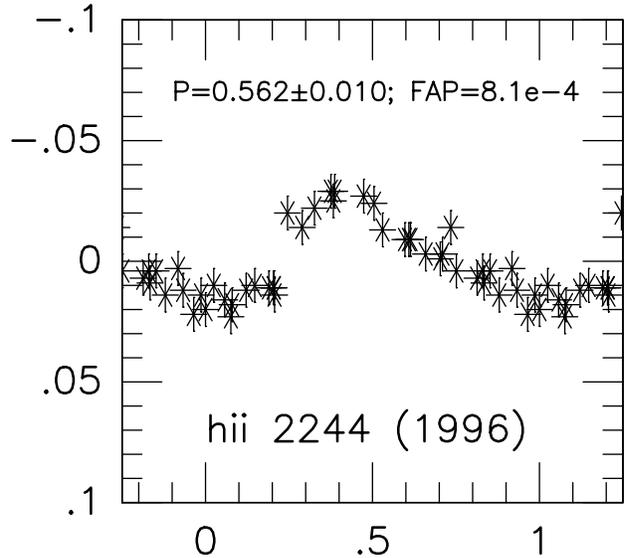
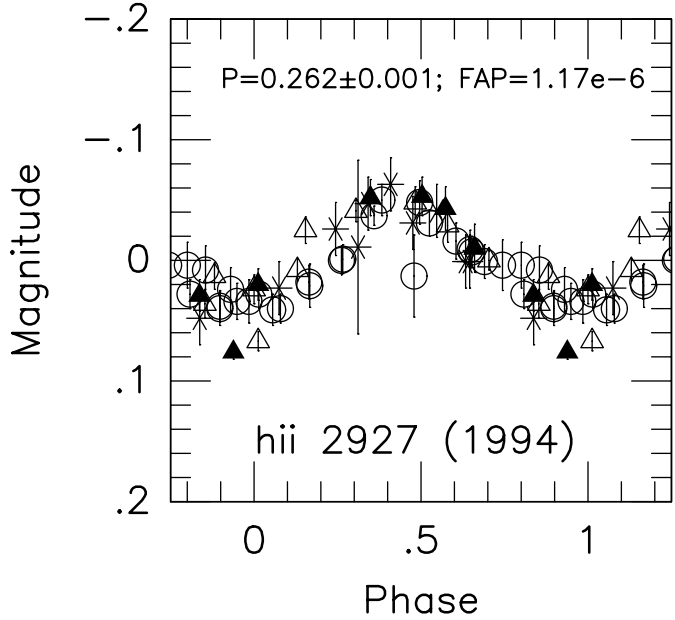
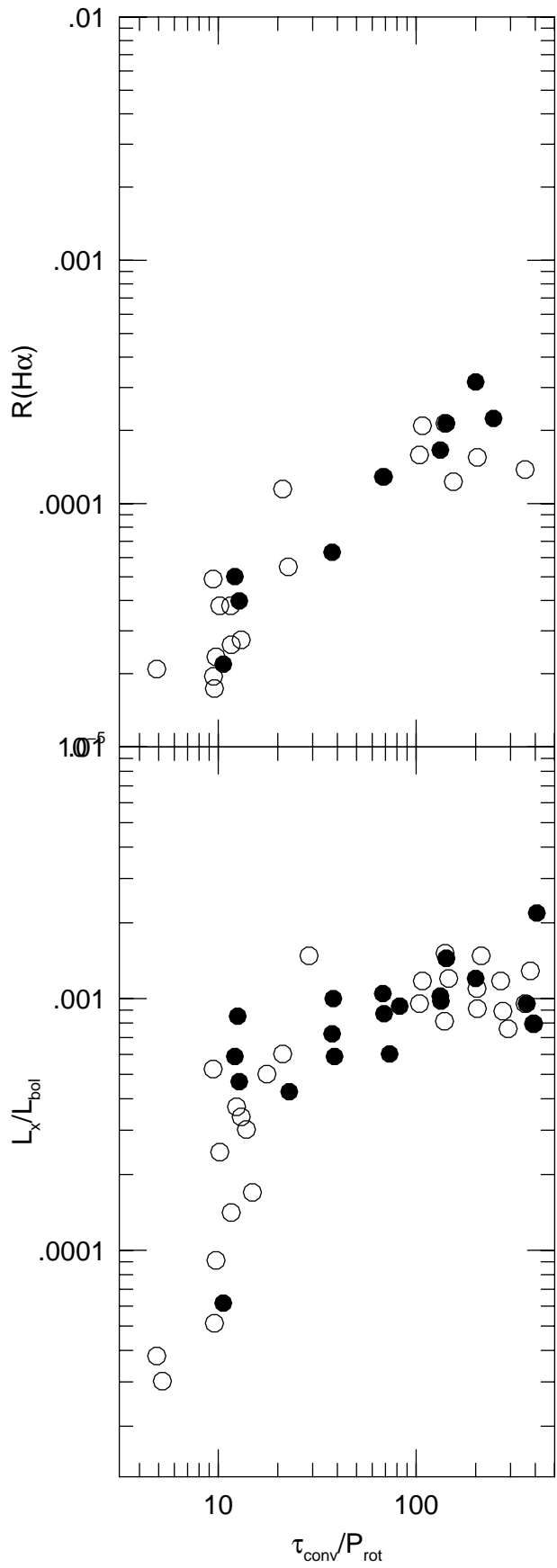
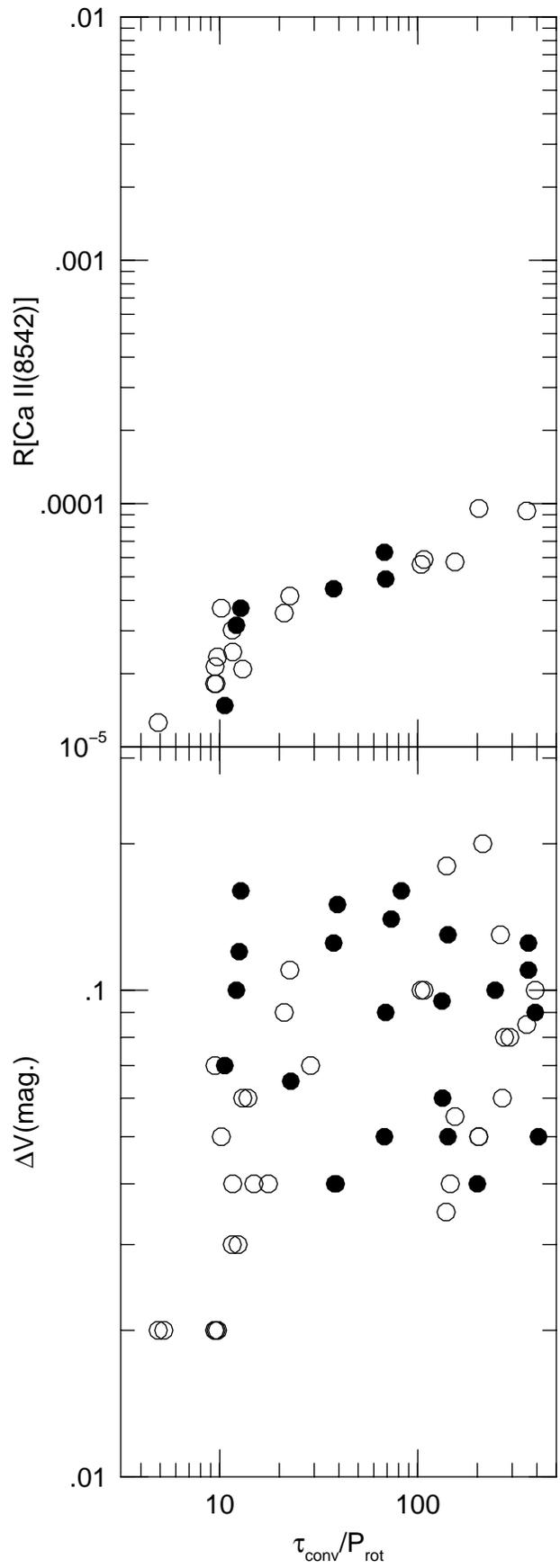
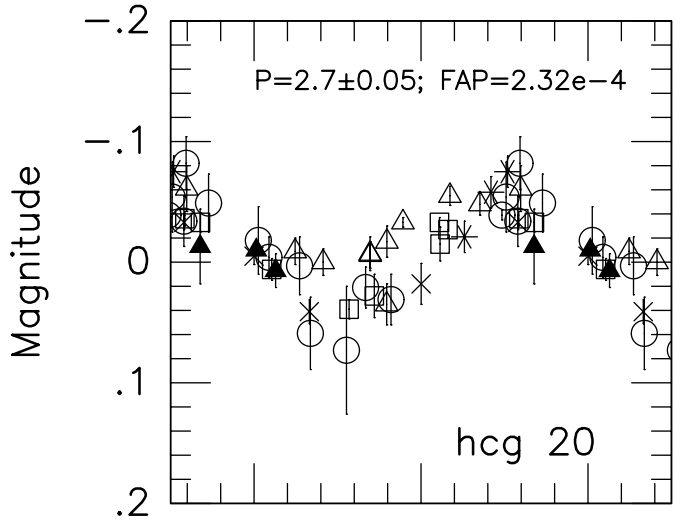
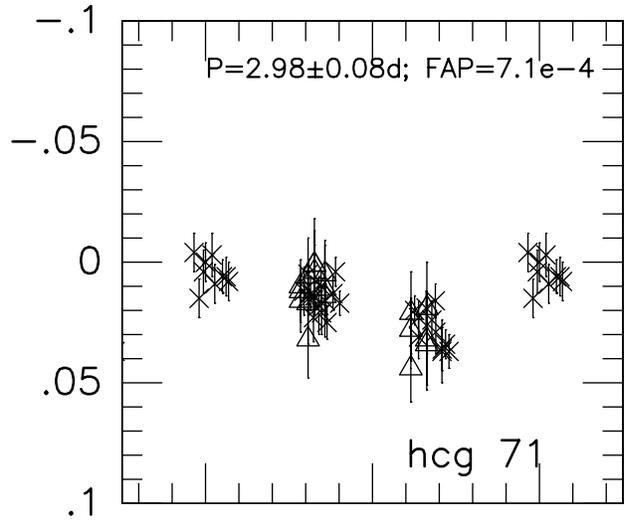
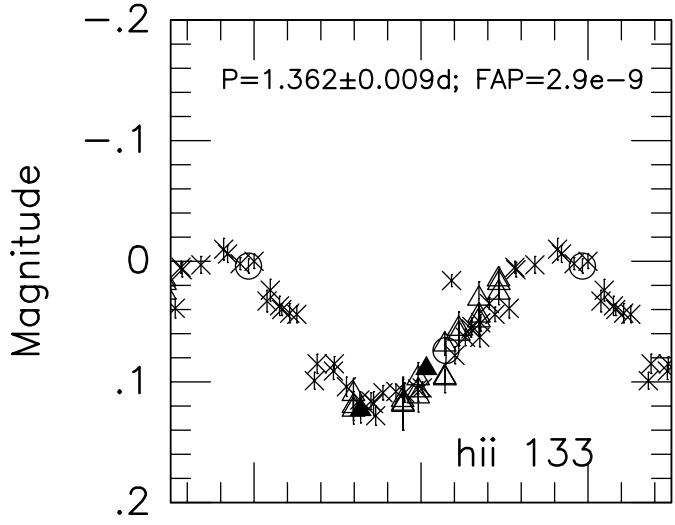
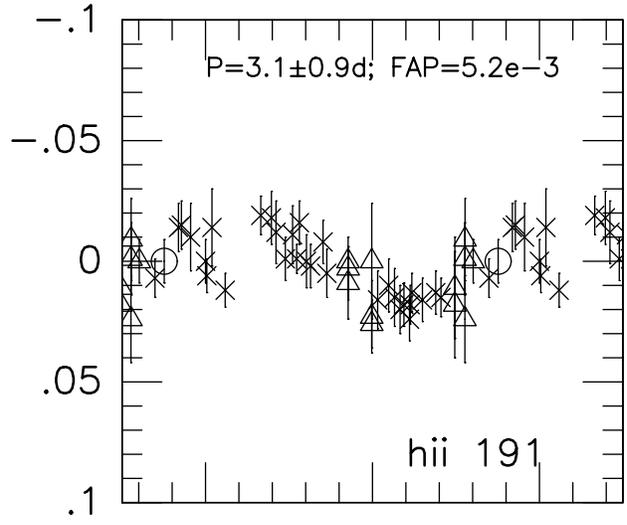
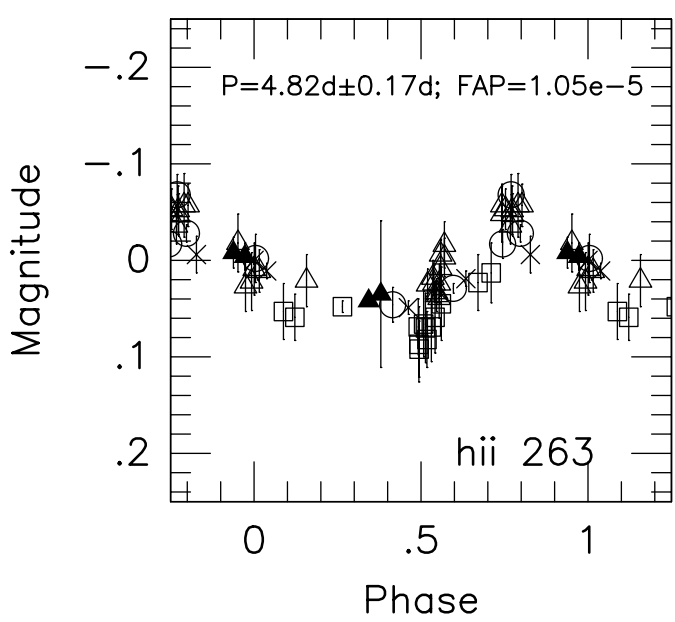
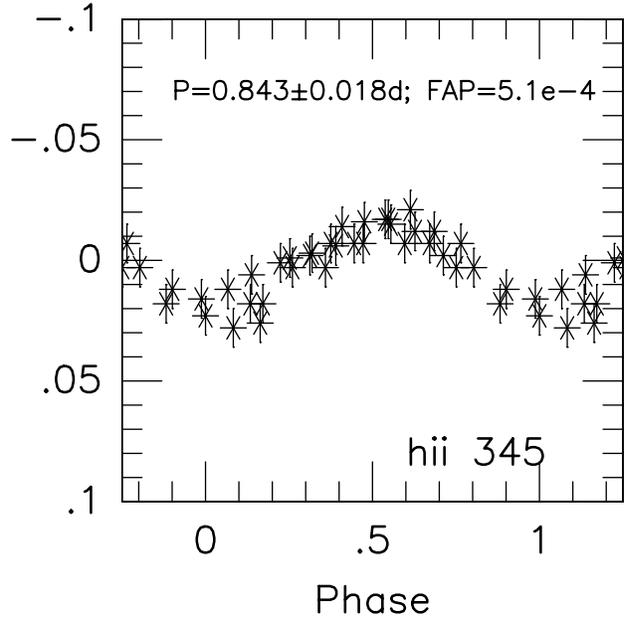
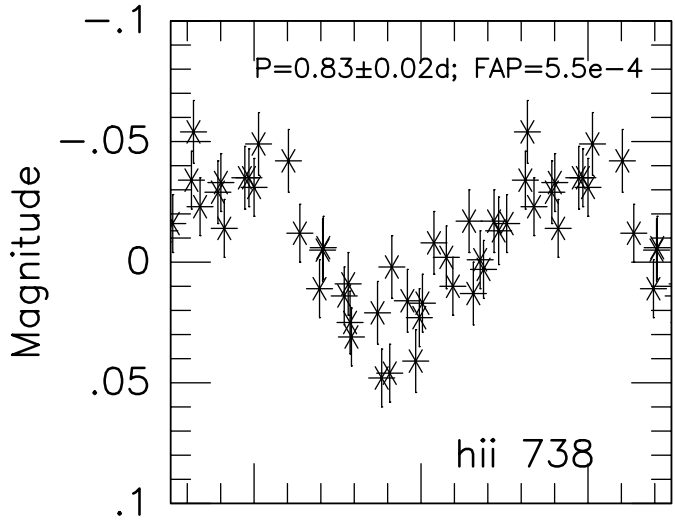
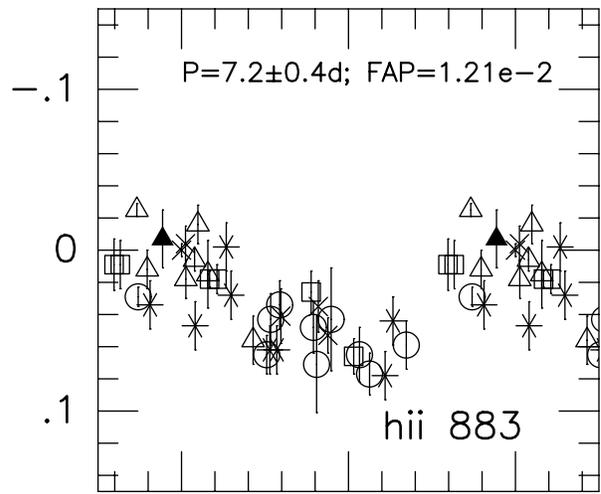
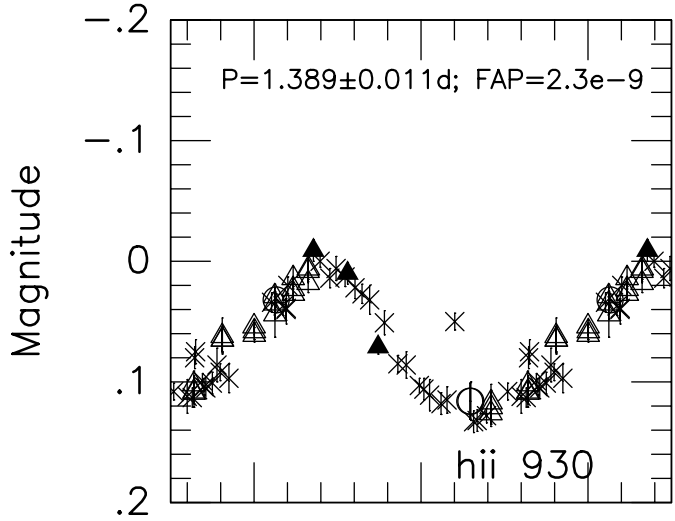
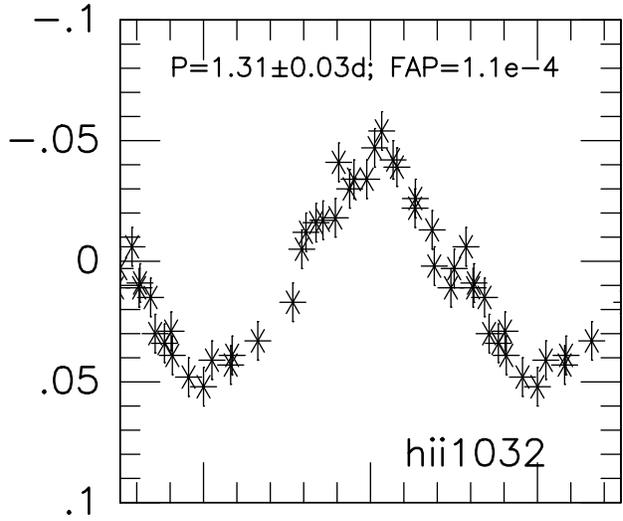
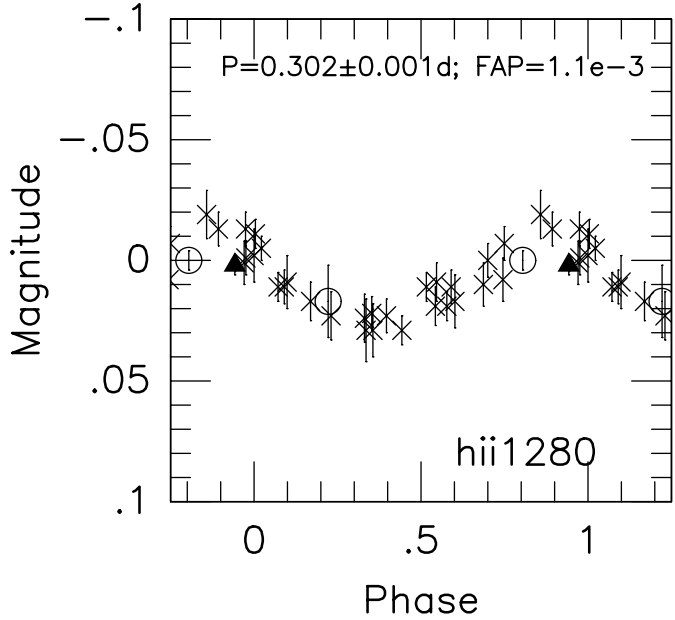
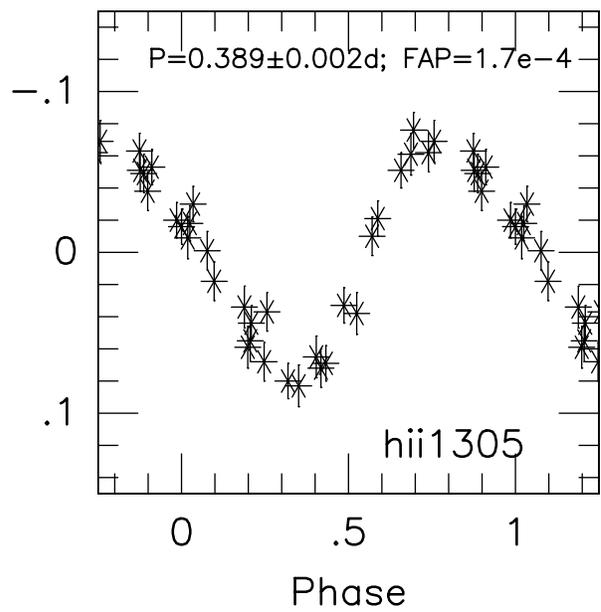
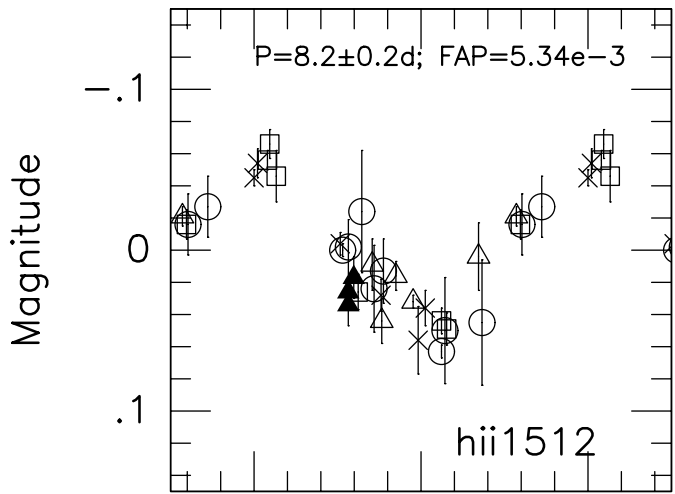
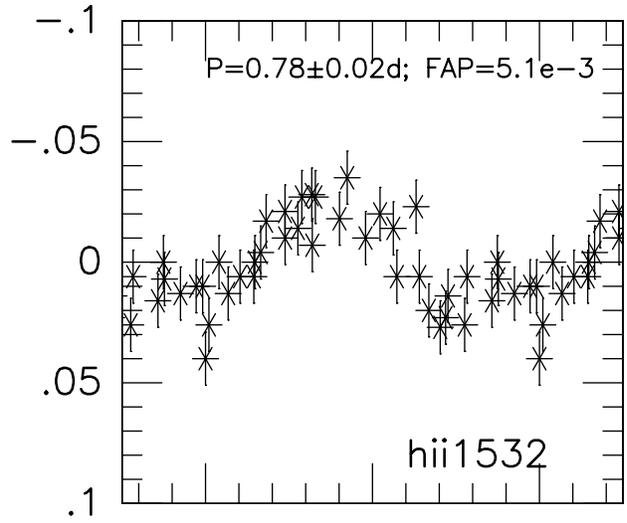
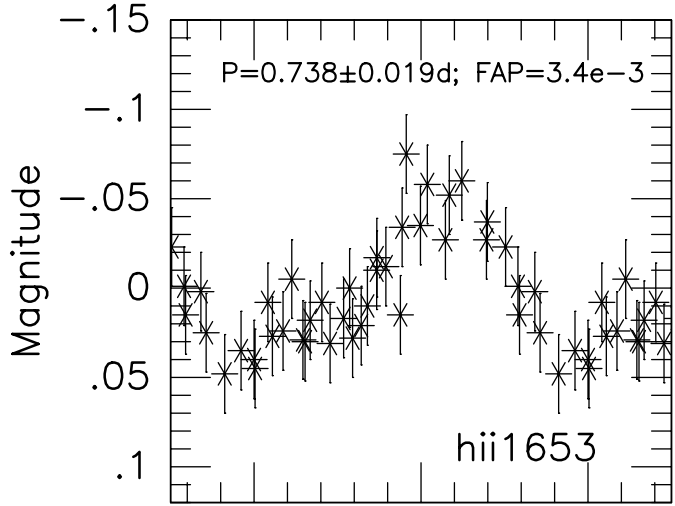
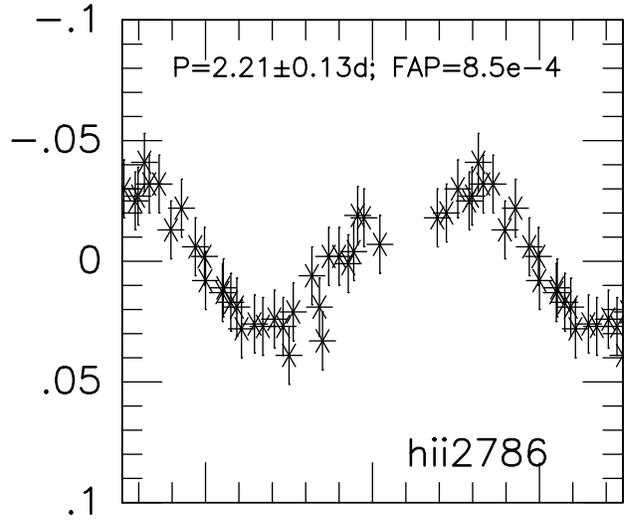
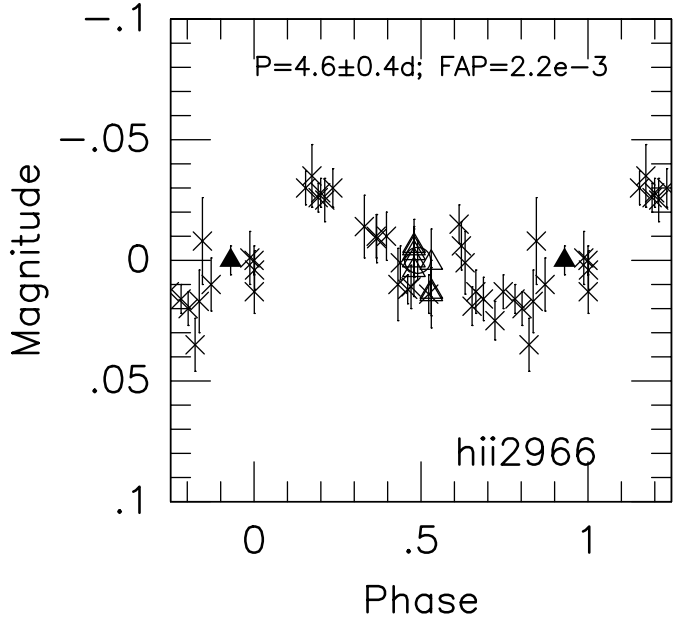
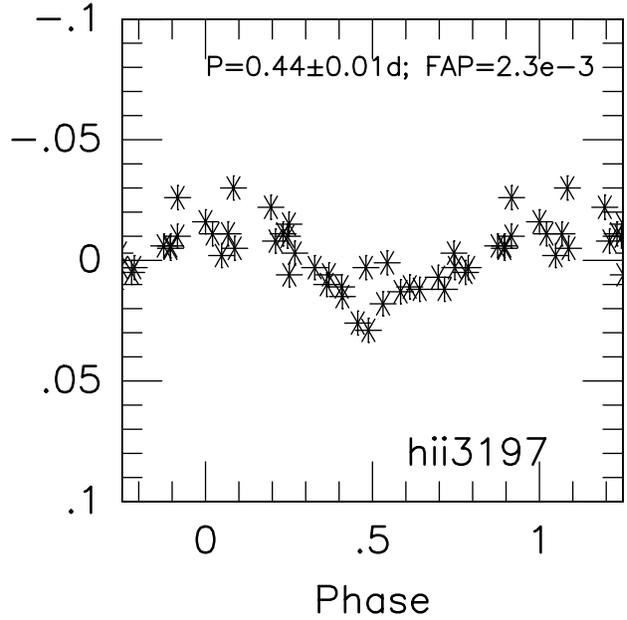
\end{document}


\begin{deluxetable}{lccc}
\tablenum{2}
\tablewidth{0pt}
\tablecaption{Summary of observations}
\tablehead{
\colhead{Star\tablenotemark{a}} &
\colhead{$N_{obs}$} & 
\colhead{Dates of obs.} & 
\colhead{Telescope\tablenotemark{b}}
} 
\startdata
Hcg20   & 40	& Oct94-Dec94 	& 1,2,3,4,5 	\\
Hcg44   & 28	& Oct94-Dec94 	& 2,3,4,5	\\
Hcg71   & 52	& Nov95-Dec95 	& 2,5 		\\
Hcg77   & 38	& Oct94-Dec94 	& 1,2,3,4,5 	\\
Hcg103  & 27	& Oct94-Dec94 	& 2,3,4,5   	\\
Hcg195  & 27	& Oct94-Dec94 	& 1,2,3,4,5 	\\
Hcg219  & 31	& Oct94-Dec94 	& 2,3,4,5   	\\
Hcg380  & 26	& Nov95-Dec95 	& 2,5       	\\
Hcg422  & 26	& Nov95-Dec95 	& 2,5 	    	\\
Hii133  & 61	& Nov95-Dec95 	& 2,4,5     	\\
Hii174  & 38	& Oct96	      	& 1	    	\\
Hii191  & 43	& Nov95-Dec95 	& 2,4,5     	\\
Hii212  & 41	& Nov95-Dec95 	& 2,5       	\\
Hii253  & 37	& Oct96	      	& 1	    	\\
Hii263  & 54	& Oct94-Dec94 	& 1,2,3,4,5 	\\
Hii345  & 34	& Oct96	      	& 1		\\
Hii708  & 37	& Oct96	      	& 1		\\  
Hii738  & 37	& Oct96	      	& 1		\\
Hii793  & 37	& Nov95-Dec95 	& 2,4,5		\\
Hii883  & 37	& Oct94-Dec94 	& 1,2,3,4,5	\\
Hii930  & 63 	& Nov95-Dec95 	& 2,4,5		\\
Hii1029 & 26	& Nov95-Dec95 	& 2,5		\\
Hii1032 & 34	& Oct96	      	& 1		\\
Hii1280 & 32	& Nov95-Dec95 	& 2,5		\\
Hii1305 & 32	& Oct96	      	& 1		\\
Hii1512 & 32	& Oct94-Dec94 	& 1,2,3,4,5 	\\
Hii1532 & 36	& Oct96	      	& 1		\\
Hii1553 & 36	& Oct96	      	& 1		\\
Hii1653 & 36	& Oct96	      	& 1		\\
Hii1883 & 47	& Oct94-Dec94 	& 1,2,3,4,5	\\
	& 38	& Oct96	      	& 1		\\
Hii2147 & 37	& Oct96	      	& 1		\\
Hii2244 & 47	& Oct94-Dec94 	& 1,2,3,4,5	\\
	& 37	& Oct96	      	& 1		\\
Hii2786 & 34	& Oct96       	& 1		\\
Hii2927 & 52	& Oct94-Dec94 	& 1,2,3,4,5	\\
Hii2966 & 43	& Nov95-Dec95 	& 2,5		\\
Hii3197 & 37	& Oct96	      	& 1				
\tablenotetext{\rm a}{Hii refers to the second list in Hertzsprung 1947,
sometimes also referred to as Hz; Hcg refers to the catalog of Haro,
Chavira \& Gonzalez 1982} 
\tablenotetext{\rm b}{1: NURO 0.8m, 2: KPNO 0.9m, 3: Lowell 1.1m, 4: SAO
1.2m, 5: Lowell 1.8m}
\enddata
\end{deluxetable}